\documentclass[letterpaper, 10 pt, conference]{ieeeconf}  

\IEEEoverridecommandlockouts                              

\usepackage{hyperref}
\usepackage{tabularx,booktabs}
\usepackage{makecell}
\usepackage{graphicx}

\title{\LARGE \bf
From Dyad to Triad: Eliciting XAI Requirements in Stroke Rehabilitation
}

\author{ Param Rajpura$^{1}$, Yogesh Kumar Meena $^{1}$
\thanks{$^{1}$ Param Rajpura and Yogesh Kumar Meena are with Human-AI Interaction (HAIx) Lab, IIT Gandhinagar, India
        {\tt\small yk.meena@iitgn.ac.in}}%
}

\def\BibTeX{{\rm B\kern-.05em{\sc i\kern-.025em b}\kern-.08em
    T\kern-.1667em\lower.7ex\hbox{E}\kern-.125emX}}

\begin{document}

\maketitle
\thispagestyle{empty}
\pagestyle{empty}

\begin{abstract}

Eliciting explainable AI (XAI) requirements from stroke survivors presents a methodological challenge with direct implications for the design of trustworthy brain-computer interfaces for rehabilitation. How can patients and caregivers articulate preferences about algorithmic transparency when they lack conceptual frameworks for explainability, and when standard elicitation approaches are structurally inadequate for users with acquired communication disorders? We present a video-based scaffolding protocol for XAI requirements elicitation, developed and piloted in a rehabilitation context. In a formative study with three stroke survivors (two with moderate-to-severe aphasia) and three caregivers, facilitators employed four scaffolding approaches alongside the videos: 1) analogical bridging mapping AI states to familiar systems, 2) projective personas depersonalising sensitive topics, 3) binary forcing reducing cognitive load, and 4) extended response time. These approaches successfully surfaced heterogeneous, sometimes conflicting XAI needs across participants. Reflexive analysis additionally revealed three systematic facilitation biases, namely, normative bias, hypothesis confirmation bias, and presence effect, where scaffolding inadvertently shaped responses. We present these as protocol risk guidelines for practitioners. Together, the protocol and guidelines constitute a reusable methodological contribution for eliciting patient-facing XAI requirements in rehabilitation, arguing that such elicitation is a necessary prerequisite for trustworthy human-machine systems design, not an optional preliminary.

\end{abstract}

\section{INTRODUCTION}

\begin{figure*}[htbp!]
  \centering
  \includegraphics[width=0.85\linewidth]{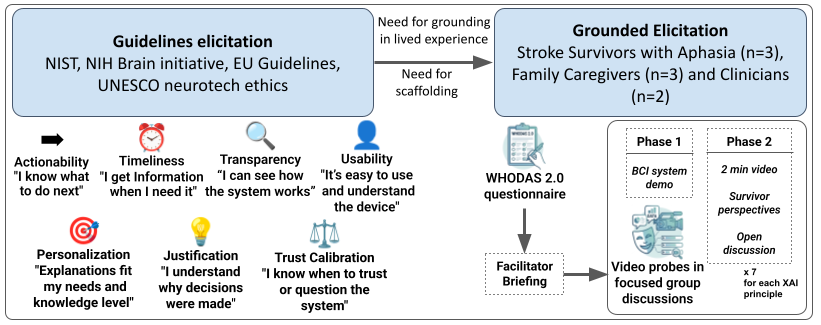}
  \caption{Activity flow of the work presented in this paper. The study with stroke survivors, caregivers, and clinicians as facilitators used video probes to elicit and understand interpretability needs. The videos were grounded on the principles of XAI proposed across multiple guideline documents.}
  \label{fig:overview}
\end{figure*}

When a stroke survivor is unable to move their hand, a therapist doesn't simply say "Keep trying." Rehabilitation evolves through moment-to-moment scaffolding-diagnosing failures, prescribing corrective actions, calibrating hope, transforming repetitive exercise into recovery~\cite{maclean2000qualitative}. As robot-assisted therapy (RAT)~\cite{veerbeek2017effects} and brain-computer interfaces (BCIs)~\cite{tonin2025} enter stroke rehabilitation, they restructure the therapeutic relationship. The clinician–patient dyad that has long anchored patient-centred rehabilitation gives way to a technology-mediated triad, in which an algorithmic agent now mediates the moment-to-moment interaction. Yet these systems do not scaffold as a therapist does. It is tempting to treat this as a gap to be closed by better algorithms, to make the therapist's explanations computable. But it is not obvious what such support should look like, for whom, or how much of it can be automated without losing what made it therapeutic. These are questions to be answered with patients, not assumed on their behalf.

In rehabilitation, explanations serve three critical functions. First, they enable self-correction. By learning to notice and fix errors (e.g., watching the affected arm in a mirror), the patient learns to troubleshoot independently, building self-efficacy, a key predictor of rehabilitation outcomes~\cite{korpershoek2011}. Second, they calibrate trust and motivation to sustain adherence across weeks of effortful practice~\cite{xing2025}. For example, in BCI-driven rehabilitation, it is essential to understand why progress is slow (due to noisy brain signals) versus why performance fluctuates (between fatigue and neuroplasticity). Third, they support shared decision-making: patients who understand system limitations can negotiate realistic goals with caregivers, preserving autonomy in chronic care~\cite{armstrong2017shared}. As rehabilitation moves toward unsupervised use, particularly critical in resource-constrained and the Global South contexts where overburdened healthcare systems make minimally supervised, low-cost BCI systems essential~\cite{pandian2018prevention}, the absence of explanations does not just reduce usability, it risks abandonment~\cite{gwynne2025adherence}. The interpretive labour that therapists perform must increasingly be facilitated by the technology itself.

Yet stroke survivors with aphasia (30\% of stroke patients~\cite{engelter2006epidemiology}), cognitive impairments or limited digital literacy, alongside untrained family caregivers, 
are excluded from AI design processes~\cite{newman2022definition}. Current XAI in healthcare focuses on clinical decision support for clinicians, but patient-facing rehabilitation requires a shift from justification to enablement and trust calibration~\cite{Rajpura_2024}. While BCI efficacy is established, explanatory feedback remains underdeveloped. Exclusion of end-users with communication barriers risks embedding bias, undermining patient agency, and widening health inequities.

Current design approaches for XAI often assume stakeholders already grasp key concepts. Specifically assuming that AI systems make decisions which can be made transparent, that explanations vary in granularity or format, and that trade-offs exist between interpretability and accuracy~\cite{miller2019explanation}. 
Linguistic fluency and a mental model of algorithmic systems are required to answer abstract questions about whether counterfactual or contrastive explanations are preferred~\cite{ribeiro2016why}. Therefore, it is required to first scaffold understanding of XAI principles through concrete experience, and then elicit requirements. While participatory design uses video probes and personas to ground discussions~\cite{newell2011user,moffatt2004participatory}, these primarily address interface design (layouts, icons). The challenge of eliciting requirements about algorithmic properties (transparency, actionability, trust calibration) from populations with cognitive and communication barriers remains unaddressed. Moreover, co-design with such populations risks what scholars term "empowerment theatre": facilitators unconsciously steering toward predetermined outcomes ~\cite{vines2013configuring,hamid2025}.
Researchers can inadvertently silence preferences by re-explaining until participants correctly choose complexity, reproducing ableist assumptions~\cite{vines2013configuring}. Recognising these risks as inevitable in clinical contexts, we treat facilitator–participant interactions as analysable data, to understand where scaffolding enables versus biases articulation.

Inclusive AI design and human-centred explainable AI (HCXAI)~\cite{hamid2025,ehsan2024} demand that stakeholders can meaningfully shape how AI in human-machine systems (HMS) explain their behaviour. Yet current design approaches for XAI  assume participants can discuss explainability abstractly. This work addresses this gap by demonstrating how video-grounded scaffolding enables nuanced articulation of XAI requirements from stroke survivors experiencing linguistic and cognitive barriers, while making facilitator influence visible through reflexive analysis. This work contributes to the literature by: (1) introducing a scaffolding protocol to elicit XAI requirements from stroke survivors experiencing linguistic and cognitive barriers; (2) offering a reflexive analysis of facilitation bias in participatory XAI elicitation with stroke survivors; and (3) deriving guidelines for balancing scaffolding and reflexivity, enabling researchers to involve users in shaping AI in HMS without imposing predetermined assumptions.

\section{Methodology}

\subsection{Study Design and Context}

We conducted a formative study at a neuro-rehabilitation centre in Central India 
(December 2025), approved by the Institutional Review Board (IITGN/IEC/2025-26/3). This matters as India has 0.36 physiotherapists per 10,000 population, and only 40\% of stroke survivors access rehabilitation services~\cite{bharati2021}. Six participants (3 stroke survivors with hemiparetic upper limb impairment and 3 co-residing family caregivers; ~\autoref{tab:participants}) were recruited through clinic referrals. Two facilitators 
(rehabilitation physician and neurologist with 6-month to 2-year therapeutic relationships) received a briefing on XAI principles and neutral prompting strategies.

\begin{table*}[!htbp]
  \caption{Participant Demographics (n=6: 3 stroke survivors, 3 caregivers). All participants were in the chronic phase of stroke recovery (0.5-2 years post-stroke). Score (WHODAS-2.0) ranges from 12 (no disability) to 60 (extreme). Communication assessed via clinical observation and caregiver report. Pseudonyms used.}
  \label{tab:participants}
  \begin{tabularx}{\textwidth}{lp{1.7cm}p{1.4cm}p{0.8cm}p{3.6cm}p{5.2cm}X}
    \toprule
    \thead{ID} & \thead{Gender/Age} & \thead{Education} & \thead{Score} & \thead{Communication Ability} & \thead{Technology Experience} & \thead{Caregiver} \\
    \midrule
    P1 & F, 61-75 & Graduate & 31 & Mild attention deficit & High: smartphone/tablet, adapts easily & C1 (Husband) \\
    P2 & M, 61-75 & Primary & 21 & Moderate receptive aphasia & Low: operates TV, open to learning & C2 (Adult son) \\
    P3 & M, 46-60 & Graduate & 18 & Severe expressive aphasia & High: laptop \& smartphone, tech anxiety & C3 (Adult son) \\
    \bottomrule
  \end{tabularx}
\end{table*}

\subsection{Materials: Video Scenario Development}

Video scenarios served as the core material for scaffolding XAI understanding within BCI-driven rehabilitation contexts ({\href{https://www.youtube.com/playlist?list=PLVqEA4Rk0wPc}{Link for Videos}}). Scenarios depicted realistic rehabilitation situations observed in ongoing clinical trials, enacted by actors for privacy protection. Seven scenarios (\autoref{tab:scenarios}) were grounded in XAI principles synthesising NIST guidelines~\cite{phillips2021four}, neuroethics frameworks~\cite{neuroethics2023,bianchi2018neuroethics}, and BCI-specific XAI research~\cite{Rajpura_2024}: transparency, justification, timeliness, actionability, personalisation, 
trust calibration, and usability. Unlike developer-facing interpretability~\cite{hsieh2023xbrainlab} or clinician-facing interfaces~\cite{kim2023designing,kim2024stakeholder}, our scenarios are centred on patient-facing explainability, where explanations must enable action and calibrate trust. All videos (~2 minutes each, filmed in Hindi) featured the same protagonist (Sunita) for narrative continuity. Each depicted a realistic dilemma without showing explicit system explanations, prompting participants to articulate what information would help rather than evaluate predetermined solutions. \autoref{fig:overview} illustrates the procedure and the activity workflow of the study.

\begin{table}[htbp]
  \caption{Seven video scenarios mapped to XAI principles and rehabilitation contexts. Each depicts a realistic situation without explicit explanations, prompting open discussion.}
  \label{tab:scenarios}
  \begin{tabularx}{\linewidth}{p{2.2cm}X}
    \toprule
    \thead{Scenario\\ \textit{(XAI Principle)}} & \thead{Situation \\ \textit{(Patient Question)}} \\
    \midrule
    Complex Interface\newline \textit{(Usability)} & User in the first session of home-rehabilitation, screen with technical terms, 65-page manual. \newline \textit{("I forgot the instructions I received yesterday. How do I even start?")} \\
    \addlinespace
    Inconsistent Success\newline \textit{(Justification)} & Success varies unpredictably. User fails in 2 attempts, then succeeds. \newline \textit{("I'm doing the same thing... why does it work sometimes and not others?")} \\
    \addlinespace
    Generic Feedback\newline \textit{(Personalisation)} & System gives the same goals to users with different needs (eating vs. typing). \newline \textit{("It tells me shoulder movement, but I just need to hold a spoon...")} \\
    \addlinespace
    Data Privacy\newline \textit{(Transparency)} & System records brain signals; patient wonders about data capture scope. \newline \textit{("Does it read only hand movements or my other thoughts? Who sees my data?")} \\
    \addlinespace
    Delayed Feedback\newline \textit{(Timeliness)} & No immediate feedback; report arrives 10 hours later. \newline \textit{("I don't remember what I did this morning. How does this help?")} \\
    \addlinespace
    Diagnostic Without Action\newline \textit{(Actionability)} & System says "weak signal" and "focus more" but no guidance on how. \newline \textit{("It tells me what's wrong, but not how to fix it. What should I do?")} \\
    \addlinespace
    Conflicting Assessments\newline \textit{(Trust Calibration)} & Doctor says "good progress," BCI shows 58/100. \newline \textit{("Doctor says I'm improving, but this score looks bad. Which should I trust?")} \\
    \bottomrule
  \end{tabularx}
\end{table}

\subsection{Procedure}

Sessions were conducted at the rehabilitation centre in December 2025, lasting approximately 70 minutes in two phases. In Phase 1 (20 minutes), Facilitator F1 introduced the study purpose, emphasising no right/wrong answers. Researchers then demonstrated a motor imagery (MI) BCI system using a 16-channel wireless EEG headset and 2-DOF upper limb exoskeleton~\cite{handexo1,handexo2,handexo3,handexo4}. The system detects left- and right-hand MI using common spatial pattern classification, providing real-time visual feedback (on-screen avatar) and mechanical feedback (exoskeleton-assisted movement). Participants tried the equipment and asked questions before viewing scenarios. In Phase 2 (40 minutes), seven scenarios (\autoref{tab:scenarios}) were presented sequentially (5-9 min each). After viewing each 2-minute video, facilitators asked: "What do you think about this situation?  What information would help Sunita?" allowing extended response time (15-30 seconds). Follow-up probes used concrete prompts (e.g., "Should the system tell why it failed?") or everyday analogies when needed.

Facilitators received a briefing on XAI principles but no scripted protocols, reflecting aphasia communication research's emphasis on responsive strategies~\cite{kagan1998supported} and our goal to document naturalistic clinical facilitator-participant dynamics~\cite{pilnick2011power}. This trade-off gains ecological validity but cannot isolate whether biases stem from clinical assumptions or incomplete XAI understanding. Caregivers were present throughout, contributing after participants or when participants needed support (60+ seconds).


\subsection{Data Collection and Analysis}

Sessions were audio-recorded with consent, transcribed verbatim in Hindi, corrected by bilingual researchers, and then translated to English. Analysis was carried out on the Hindi transcripts, so that coding was grounded in the original wording. Researchers maintained field notes documenting nonverbal communication, adapted strategies, and interjections from facilitators/caregivers. 

We employed reflexive thematic analysis~\cite{Braun01012006}. The first author inductively coded transcripts, identifying segments where participants articulated needs, expressed confusion, or responded to scaffolding. Initial codes included privacy concerns, trust preferences, granularity needs, use of analogy, and binary choice responses. Through iterative team discussion, to interrogate and enrich the analysis and surface alternative readings, codes were developed and organised. Three themes were generated: (a) scaffolding approaches, (b) divergent XAI preferences, and (c) facilitation dynamics. Field notes contextualised non-verbal cues. All authors then reflexively examined facilitator moves that shaped responses, identifying expert override, hypothesis confirmation, and authority effects, a critical reflexive step given power dynamics in clinical settings~\cite{Olmos-Vega04032023,Hamraie2019Crip}. Patient-only, caregiver-mediated, and jointly produced preferences were distinguished during coding and reported separately, so that each preference's source remains clear when discussing individual participants' XAI needs.

\section{Findings}
The protocol successfully elicited complex XAI requirements from stroke survivors with aphasia, demonstrating that structured scaffolding can surface nuanced, heterogeneous preferences that standard elicitation approaches would likely miss. We organise findings around two themes: the scaffolding approaches that enabled articulation, and the divergent XAI preferences revealed during the discussions.

\subsection{Scaffolding Approaches That Enabled Articulation}

Four scaffolding techniques proved effective in bridging the gap between abstract XAI concepts and participants' rehabilitation experiences. Each is presented with evidence of what it unlocked that direct questioning fails to surface.

\textit{Analogical bridging:}
Facilitators translated invisible processes into familiar systems with observable failure modes, most effectively in Scenarios 5 and 6. 
(i)~\textit{Mobile Tower Analogy (Actionability):} When discussing signal 
quality errors, F1 introduced the mobile phone signal analogy, which immediately resonated with participants who had experienced dropped calls. P2, who had remained non-responsive to direct technical questioning, responded by demanding diagnostic guidance: ``Increase the signal quality, tell me what mistake I'm making.'' The metaphor made EEG impedance tangible as a solvable problem rather than an abstract technical failure. 
(ii)~\textit{School Report Card Analogy (Timeliness):} When participants struggled to articulate desired feedback granularity, F1 used the school report analogy: " So, just like in school, each subject's score is provided, would you prefer a score for each exercise? P2 immediately requested a per-task breakdown. This demonstrated that the barrier was not the absence of preference but the absence of a conceptual anchor, the gap fulfilled by an analogy.
(iii)~\textit{Love/Secrets Metaphor (Transparency):} To discuss privacy boundaries, F1 used an emotional metaphor about personal secrets, helping participants conceptualise data minimisation as setting appropriate 
personal boundaries rather than as a technical constraint (S4-08, S4-12). This strategy unlocked the articulation of privacy preferences, where direct data-framing had produced silence.

\textit{Projective personas:}
Direct questions about sensitive topics often triggered silence or socially desirable answers. Third-person framing systematically unlocked authentic concerns, functioning as a depersonalisation device that reduced 
social desirability pressure.
(i) "Sceptical People" Persona (Transparency): When F1 asked participants directly about data privacy, P1 and P3  remained silent. Reframing as ``Some sceptical people might worry about this...'' prompted caregivers C1 and C2 to admit fears and elicited P3's opinion, surfacing privacy concerns that direct questioning had suppressed. 
(ii) Persona Scaffolding (Personalisation): The ``Home Guy vs. Office Guy'' personas helped participants understand personalisation, shifting P2 from ``Exercises should be same'' to accepting differential treatment. However, P2's underlying fairness reasoning was never probed, revealing a facilitation gap.

\textit{Binary forcing:}
Reducing complex trade-offs to A/B choices enabled clear articulation. 
(i) Machine vs. Human Mediation (Actionability): The probe ``Should machine tell you or should a human explain?'' revealed a critical design split: P1 and P2 chose machine autonomy while P3 chose human mediation. This divergence wouldn't have surfaced through open-ended questions on actionability.
(ii) Small vs. Big Report (Timeliness): ``Do you want a small summary or a big report with details?'' produced clear preferences: P1 chose small, P2 chose big. 
(iii) Persistence vs. Explanation (Justification): When faced with repeated failures, 4 out of 5 participants chose explanation over blind persistence, with P2 providing an unprompted rationale.

\textit{Extended response time:}
Facilitators consistently allowed 15-30 seconds of silence after questions. This was particularly critical for participants with expressive aphasia, enabling them to formulate responses without time pressure, a key principle in stroke communication rehabilitation.

\subsection{Divergent XAI Preferences}
\label{divergentprefs}

A key validation of the protocol's sensitivity is that it revealed stroke survivors to be a heterogeneous user group with divergent, sometimes conflicting XAI needs. Such divergent needs would be invisible to an elicitation approach, which does not penetrate communication barriers or social desirability effects. Three critical divergences emerged:

\textit{Granularity:}
Scenario 5 (Timeliness) revealed a clear split in the desired level of information detail. P1 stated a preference for minimal information (``want small report'', stated twice), while P2 demanded granular per-task breakdowns. The divergence itself is the methodological signal. An open-ended situation via the video, followed by binary forcing, surfaced both ends of this spectrum. Not eliciting this would produce misleading design requirements.

\textit{Trust:}
Scenario 7 revealed divergence in epistemic trust (who participants believe is correct). P1 and P3 deferred to human authority. When asked about conflicting assessments between the clinician and BCI scores, P1 trusted the doctor. P2, by contrast, expressed trust in quantified machine feedback, viewing numbers as more objective than qualitative clinical assessment. Notably, behavioural trust converged regardless of epistemic preference. Through sustained probing, all participants stated they would ``work harder'' irrespective of which source they believed. This suggests that trust conflicts may not paralyse rehabilitation adherence, a critical insight that needs to be verified at scale for deployment of human-machine systems where clinicians aren't present.

\textit{Agency:}
Scenario 6 exposed preferences for error mediation that diverged along lines invisible without persona and binary scaffolding. P1 and P2 wanted direct machine guidance for autonomous self-correction. However, P3 preferred family-mediated explanations, revealing lower self-efficacy with technical systems. Moreover, this resonated with P3's opinion on their shift in technology experience post-stroke, as reported in the questionnaire (\autoref{tab:participants}). This split reflects differences in technology anxiety and caregiver availability, both of which are critical factors in the design of home/clinic rehabilitation systems.

\section{Discussion}

This study makes two arguments relevant to the BCI and trustworthy human-machine systems (HMS) community. First, patients are legitimate XAI stakeholders whose needs are distinct from, and not reliably proxied by, clinicians~\cite{valentine2022}. Second, requirements elicitation from patients is a necessary step, not an optional preliminary, for the BCI development pipeline aimed at translation beyond lab premises. Building on the protocol's risks and guidelines, we discuss each argument in turn, followed by an account of the limitations and future scope.

\subsection{Protocol Risks as Methodological Guidelines}
\label{protocolrisks}
Reflexive analysis exposed three systematic risks in the protocol's execution. We present these not as failures but as derived guidelines that any practitioner replicating or adapting the XAI requirements elicitation protocol must explicitly guard against. 

\textit{Normative bias (expert override of minority preferences):}
In Scenario 5, P1 stated ``want small report'' twice, yet F2 summarised ``they want full report'', erasing this preference. This bias unconsciously values complexity because it aligns with the capabilities of AI systems. The bias is structurally predictable when facilitators have technical expertise. It treats simplicity as a failure of comprehension rather than a valid design requirement. \textit{Guideline: facilitators 
must implement explicit minority-preference protection, recording and probing divergent responses before synthesising.}

\textit{Hypothesis confirmation bias:}
When P2 expressed a pro-surveillance stance (``Good thing a machine should know other thoughts''), the facilitator immediately pivoted to elicit the expected privacy-fear response from another participant. 
Counter-intuitive responses, precisely those most valuable for robust requirements, were the most vulnerable to being bypassed. \textit{Guideline: counter-intuitive responses must trigger deeper probing, not redirection.}

\textit{Presence effect:}
Medical authority in the room biased trust responses in Scenario 7. Notably, F1 identified and named this bias: ``If I wasn't here, you'd say machine.'' This reflexive correction is a methodological contribution, as it demonstrates that facilitator reflexivity can partially recover from presence effects in real time and validates the need for such acknowledgement and for selectively choosing specific sections of the discussions to be anonymous or peer-only elicitation conditions in future protocol iterations. \textit{Guideline: Preference elicitation on authority-sensitive topics needs to acknowledge the power dynamics or avoid it without hampering the natural clinician-patient interaction scenarios.}

\subsection{Patients as Legitimate XAI Stakeholders}

The divergences reported in ~\autoref{divergentprefs} provide a direct empirical case that patients hold distinct, non-trivial XAI needs: preferences over granularity, epistemic trust, and error mediation diverged, and sometimes conflicted, across just three survivors. These are genuine disagreements about what a trustworthy system should do. The behavioural-trust finding sharpens why patients must co-design rather than be proxied. Despite diverging on whom they trusted, all participants converged on the same behavioural response: work harder. Patient-facing XAI, therefore, need not resolve epistemic disagreements about authority to support adherence; it must support engagement, agency, and interpretability at the patient's level, where the goals are distinct from clinical decision support and require separate design attention.

\subsection{Elicitation as a Development Prerequisite}

This study demonstrates that direct elicitation of XAI requirements from stroke survivors is both feasible and necessary. The techniques described in the findings are not complex; they are careful. Analogical bridging, projective personas, binary forcing, and extended response time are individually known in communication research~\cite{casarett2010,hurde2025,kagan1998supported,ryan2004}. Their systematic combination, alongside video materials grounded in XAI principles, acting as an XAI elicitation protocol, is the core contribution.

The facilitation risks exposed through reflexive analysis (\autoref{protocolrisks}) establish the necessity of such elicitation. These are not idiosyncratic failures of particular facilitators. They are structurally predictable when technically expert researchers elicit requirements from patients in clinical settings. They would operate silently in any informal or 
unexamined elicitation process, producing design inputs that reflect the researcher's assumptions more than the patient's needs. A BCI system built on such inputs may be technically transparent yet fail to support meaningful patient engagement, leading to a trustworthiness failure that originates before technical development.

\subsection{Limitations and Future Directions}

The primary contribution of this work is methodological. The scaffolding protocol and the facilitation risk guidelines are the transferable outputs, not the specific XAI preferences of this participant group. The sample size (N=6) is sufficient to demonstrate that the method surfaces heterogeneous requirements and to derive protocol guidelines, but not sufficient to characterise what stroke survivors want from XAI systems as a population, and we make no such claim.

The protocol was developed and piloted in a single rehabilitation centre with specific socio-economic, linguistic, and clinical resource contexts. Whether the same scaffolding techniques prove effective in different rehabilitation settings, with different aphasia profiles, or with non-clinical BCI users remains an open question. Scenarios were grounded in MI BCI with exoskeleton feedback; other paradigms may surface different explainability needs and require different analogical anchors. Finally, this study documents stated preferences elicited under controlled conditions, not behaviour in deployed systems. Whether requirements gathered through this protocol produce better-aligned patient-facing XAI than clinician-proxied design assumptions is the critical empirical question this work motivates but cannot yet answer.

\section{Conclusion}


This work proposes a video-based scaffolding method for eliciting patient-facing XAI requirements from stroke survivors with moderate-to-severe aphasia. Across seven BCI-driven rehabilitation scenarios, facilitation approaches such as analogies, projective personas, binary forcing, and extended response time helped participants articulate nuanced needs that were difficult to surface in abstract discussion. However, reflexive analysis shows that scaffolding can also shape responses through expert override, hypothesis confirmation, and authority effects. Beyond BCI rehabilitation, the method extends to other patient- or caregiver-facing AI settings with similar communication barriers and clinical authority. It thereby realises one stage of a broader agenda~\cite{rajpura2026} for co-designing and evaluating patient-facing XAI in BCI-driven stroke rehabilitation.

\textbf{Acknowledgement:} We gratefully acknowledge Apoorva Pauranik and Ankit Sodani for facilitating the study, Girijesh Prasad and Saugat Bhattacharya for providing the physical prototype of the exoskeleton, Hina Rajpura for her support during video production, and Khondaker A. Mamun for supporting the IBRO Fellowship. This work was supported by the IBRO Exchange Fellowship (EF-0274153858) and the IIT Gandhinagar startup grant (IP/IITGN/CSE/YM/2324/05).

\bibliography{SMC_Scaffolding}

\end{document}